\documentstyle[preprint,aps]{revtex}
\draft
\title
{\bf \LARGE{PERSISTENT CURRENTS IN THE PRESENCE OF A
TRANSPORT CURRENT}}
\author{A. M. Jayannavar and P. Singha Deo }
\address{Institute of Physics, Sachivalaya Marg,
Bhubaneswar 751005, India}
\begin{document}
\maketitle
\begin{abstract}
We have considered a system of a metallic ring coupled to two
electron reservoirs. We show that in the presence of a transport
current, the persistent current can flow in a ring, even in the
absence of magnetic field. This is purely a quantum effect and
is related to the current magnification in the loop.
These persistent currents can be observed if one tunes the Fermi
energy near the antiresonances of the total transmission
coefficient or the two port conductance.

\end{abstract}
\narrowtext
\newpage

Experimental and theoretical
research in mesoscopic systems have provided an opportunity of
exploring truely quantum mechanical effects beyond the atomic
realm[1]. Persistent currents in small metal rings threaded by
magnetic flux are a manifestation of quantum effects in
submicron systems and are prominent amongst the mesoscopic
effects. Prior to the experimental observations[2-4],
B$\ddot u$ttiker, Imry
and Landauer in their work suggested the existence of persistent
currents in an ordered one dimensional ring threaded by a
magnetic flux[5]. The coherent wavefunctions extending over the
whole circumference of the loop leads to a periodic persistent
current.
General quantum mechanical principles require
that the wave functions, eigenvalues and hence all observables
be periodic in a flux $\phi$ threaded by the loop with a period
$\phi_{0}$, $\phi_{0}=hc/e$ being the elementary flux quantum.
The magnetic field destroys the time reversal symmetry and as a
consequence the degeneracy of the states carrying current
clockwise and anticlockwise, is lifted. Depending on the position
of the Fermi level, uncompensated current flows in either of the
directions(diamagnetic or paramagnetic). For an ideal isolated
ring without impurities and
at zero temperature the nature of the persistent current depends
on the total number N of the electrons and the persistent
current exhibits a saw tooth type behavior as a function of
magnetic flux. For N even, the jump discontinuities occur from the
values -(2e$v_{f}$/L) to (2e$v_{f}$/L) at $\phi$=0, $\pm
\phi_{0}, \pm 2\phi_{0}$ and at $\phi$=$\pm
\phi_{0}/2, \pm 3\phi_{0}/2$ etc, for N odd. Here $v_{f}$ is
the Fermi velocity and L is the circumference of the ring.
Studies have been extended to include multichannel rings,
disorder, spin-orbit coupling and electron-electron interaction
effects[6-12]. The persistent current which flows without dissipation
is an equilibrium property of the ring and is given by flux
derivative of the total energy of the ring. These currents can
also be thought to arise from the competing requirements of
minimising the free energy in the presence of flux and at the
same time maintaining the single valuedness of the wave
function. Persistent currents are truely mesoscopic effects in
the sense that they are strongly suppressed when the ring size
exceeds the characteristic dephasing length of the electrons
$L_{\phi}$ (i.e., length scale over which the electron can be
considered to be in a pure state).

Theoretical treatments to date have been mostly concentrated on
isolated rings. Persistent currents occur not only in the
isolated rings but also in rings connected via leads to electron
reservoirs, namely in open systems[13-19]. In the recent experiment by
Mailly et. al, have measured the persistent current in both
closed and open rings[4]. B$\ddot u$ttiker gave a first conceptually
simple approach of a small metal loop connected to an electron
reservoir (open system)[13]. The reservoir acts as a source and a
sink for electrons and is characterized by a well defined
chemical potential $\mu$, and by definition there is no phase
relationship between the absorbed and emitted electrons by the
reservoir. The
reservoir acts as an inelastic scatterer and as a source of
energy dissipation or irreversibility.
All the scattering processes in the leads are assumed to be
elastic. Inelastic processes occur only in the reservoir, and
hence there is a complete spatial separation between elastic and
inelastic processes. Due to the presence of inelastic scattering
(by definition) in open systems the amplitude of persistent
current is smaller as compared to the closed systems. Weak
inelastic scattering does not destroy the effect leading to
persistent currents. We have
extended B$\ddot u$ttikers discussions to a case wherein electrons from
the reservoir enter and leave the ring in a subbarrier regime
characterized by evanescent modes throughout the circumference
of the loop[17]. In this situation the persistent current arises
simultaneously due
to two nonclassical effects, namely, Aharonov-Bohm effect and
quantum tunneling. The dependence of the current on the length
of the ring is similar to that arising due to states localized
by static disorder. In our recent work we have calculated the
persistent currents in a normal metal loop connected to two
electron reservoirs in the presence of magnetic flux[18]. We have
shown that in general the magnitude of persistent current in a
loop depends on the direction of current flow from one reservoir
to the other. Persistent currents in open systems are
sensitive to the direction of current, unlike the physical
quantities such as conductance. We hope that this effect is
useful for separating persistent current from other parasital
currents(noise) associated with experimental measurements.

In our present work, we have considered a metallic loop coupled
to two
electron reservoirs (characterized by chemical potentials
$\mu_{1}$ and $\mu_{2}$) via ideal wires as shown in fig. (1).
For the sake of simplicity we have restricted to the case of one
dimensional structure. Length of the upper arm of of the loop is
$l_{1}$ and that of the lower arm is $l_{2}$ such that the
circumference of the ring is L=$l_{1}+l_{2}$. When the chemical
potential $\mu_{1}$ is greater than $\mu_{2}$, the net current
flows from left to right and vice versa when $\mu_{2}$ is
greater than $\mu_{1}$. We show that in the presence of a
current flow through the sample (non-equilibrium situation), a
net circulating current current flows in a loop in the absence
of magnetic field in certain range of Fermi energies. In
a sense the persistent current is induced by incident carriers.
Existence of such currents were first discussed by B$\ddot u$ttiker[14],
however, our analysis is qualitatively different from that of an
earlier study. The current injected by the reservoir into the
lead around the small energy interval dE is given by $dI_{in}$=
ev(dn/dE)f(E)dE. Here v=$\hbar$ k/m is the velocity of the
carriers at the energy E, (dn/dE)=1/(2$\pi \hbar$v) is the
density of states in the perfect wire and f(E) is the Fermi
distribution. The total current flow I in a small energy
interval dE through the system is given
by the current injected into the leads by reservoirs multiplied
by the transmission coefficient T. This current splits into
$I_{1}$ and $I_{2}$ in the upper and the lower arms respectively at
the junction, such that  I=$I_{1}+I_{2}$ (conservation of
current or Kirchoff's law). Since the upper and lower arm
lengths are unequal,
in general these two currents differ in magnitude. B$\ddot u$ttiker[14],
suggests a picture for this difference as arising due to a
circulating current $I_{0}$, such that the current in the upper
branch is then given by $I_{1}=I/2+I_{0}$ and current in the
lower branch is given by $I_{2}=I/2-I_{0}$. Such a construction
always results in a persistent current. However, if this
definition is taken seriously then even in a classical loop with
different resistances in different arms one gets different
currents in the presence of a dc current
and hence persistent current. It is clear then that
with this definition one can get persistent currents even
classically without invoking quantum mechanics at all. In our
present quantum problem when one calculates the currents
($I_{1},I_{2}$) in two loops there exists two distinct
possibilities. The first possibility being for a certain range
of incident Fermi wave vectors (or energies) the current in the
two arms $I_{1}$ and $I_{2}$ are individually less than the total
currentI, such that I=$I_{1}+I_{2}$. In such a situation both
currents in two arms flow in the direction of applied field. In
such a situation we do not assign any persistent current flowing
in the ring. However, in certain energy interval, it turns out
that current in one arm is larger than the total current I
(magnification property). This implies that to conserve the
total current at the junctions, the current in the other arm
must be negative, or should flow against the applied external
field induced by difference in the chemical potentials. In such a
situation one can interpret that the negative current flow in
one arm of the loop continues to flow in a loop as a circulating
(or persistent) current. Thus the magnitude of persistent
current is the same as that of the negative current.
The direction of the persistent current can be inferred as
follows. Consider a case when the net current flows in the right
direction (i.e., $\mu_{1} > \mu_{2}$). If for this case negative
current flows in the lower arm then persistent current flows in
clockwise (or positive) direction. If on the other hand negative
current flows in the upper arm then the persistent current flows
in a anticlockwise (or negative) direction.
The negative current in one arm of the
loop is purely a quantum mechanical effect. Our procedure of
assigning persistent current, only when negative current flows
in one of the arms is exactly the same procedure well known in
classical a.c. network analysis[20]. It is well known that, when
a parallel resonant circuit (capacitance C connected in parallel
with combination of inductance L and resistance R) is driven by
external electromotive force (generator), the circulating
current arises
in LCR circuit at a resonance frequency. This effect is
sometimes referred to as a current magnification. In this
classical network when the external driving frequency is around
a resonance frequency circulating currents are possible.
Moreover at the resonance the total net current amplitude in the
circuit is at its minimum value. It turns out that even in our
quantum problem the circulating current arises near the
antiresonances (or transmission zeros) of the loop structure
coupled to leads.

We now consider a case where the current is injected from the
left reservoir (i.e., current flow is in the right direction). The
total current flow around a small energy interval is given by
I=(e/2$\pi\hbar$)T, where T is the total transmission
coefficient. It is a straight forward exercise to set up a
scattering problem for this case and to calculate the
transmission coefficient and the currents in the upper ($I_{1}$)
and the lower ($I_{2}$) arms. We closely follow our earlier method
of quantum waveguide transport on networks[17,18,21-22]
to calculate these
quantities. We have imposed the Griffiths boundary conditions
(conservation of current) and single valuedness of the
wavefunctions at the junctions. For details see
ref[17,18,21-23]. The
expressions for I, T, $I_{1}$, $I_{2}$ are given by

\begin{equation}
I=(e/2\pi\hbar)T
\end{equation}

\begin{equation}
T=(8   (2 - cos[2 k l_{1}] -
     cos[2 k l_{2}] + 4 sin[k l_{1}] sin[k l_{2}]))/\Omega
\end{equation}

\begin{equation}
I_{1}=(e/2\pi\hbar)8(1 - cos[2 k l_{2}] + 2 sin[k l_{1}] sin[k l_{2}])/\Omega
\end{equation}

\begin{equation}
I_{2}=(e/2\pi\hbar)8(1 - cos[2 k l_{1}]  + 2 sin[k l_{1}] sin[k l_{2}])/\Omega
\end{equation}

\noindent where

 $$\Omega= (37 - 5   cos[2 k l_{1}] -
    32 cos[k l_{1}] cos[k l_{2}] - 5 cos[2 k l_{2}] +
 $$
\begin{equation}
5 cos[2 k l_{1}] cos[2 k l_{2}] + 48 sin[k l_{1}] sin[k l_{2}] -
    4  sin[2 k l_{1}] sin[2 k l_{2}])
\end{equation}

\noindent Here k is the incident
wave vector. Our expression for the transmission coefficient
agrees with the earlier known expression[23] for the case of
$l_{1}$ = $l_{2}$. The transmission coefficient across a
metallic loop connected to two reservoirs and in the presence of
magnetic flux has been investigated by several authors[24,25] in
connection with the Aharonov Bohm effect.
We have first studied the behavior of the
currents $I_{1}$ and $I_{2}$ as a function of the Fermi
wavevectors. We then identify the wavevector intervals, wherein
either $I_{1}$ or $I_{2}$ flows in the negative direction and by
knowing their magnitudes we have calculated the persistent
currents as described in the earlier paragraphs.
In fig. (2) we have plotted the
circulating currents (solid curves) in the dimensionless units
($I_{c} \equiv 2\pi\hbar I_{c}/e$) in the small energy interval
dE around the Fermi energy as a function of dimensionless wave
vector kL. We have taken $l_{1}/l_{2}$=5.0/3.0. In fig. (2) we have
also plotted the transmission coefficient T for the same parameter
values. We notice that the persistent current changes sign as we
cross the energy or the wave vector at the first antiresonance
(transmission zero or minimum) in the transmission coefficient.
It does not change the sign as we cross the second
antiresonance. The first antiresonance is characterized by a
asymmetric zero-pole in the transmission amplitude (zero occurs at a
value of kL=(2$\pi$) and poles are given by kL=
(6.25495-i 0.299976) and (6.46865-i 1.90045)). The proximity of the
zero and the pole lead to the sharp variations in the
transmission coefficient around the magnitude zero as a function
of energy and lead to a asymmetrical behavior in the
transmission coefficient (around antiresonance), sometimes
termed as a Fano resonance[26]. The second antiresonance is
characterized by a zero along with symmetrically placed two
poles and the transmission coefficient is symmetric around the
antiresonance. The zero is at a value kL=(4$\pi$) and poles are given
by kL=(12.4105-i 1.07584) and (12.7222-i 1.07584).
We have thus shown that the persistent current
arises near the vicinity of the antiresonances and the nature of
the persistent current as we cross the antiresonance depends on
the zero-pole structure in the transmission amplitude around the
antiresonance.

In fig. (3) we have plotted the persistent currents in
dimensionless units (solid curves) and transmission coefficient
(dashed curves) versus kL for a case when $l_{1}/l_{2}$=(3.0). For
this particular case the transmission coefficient is symmetric
around the antiresonances and persistent current does not change
the sign as we cross the antiresonance. In general the zero-pole
structure in the transmission coefficient is sensitive to the
ratio $l_{1}/l_{2}$, being commensurate or not. For incommensurate
ratio we mostly obtain the Fano type antiresonances. For
commensurate case depending on the degree of commensuration we
can have both Fano type as well as symmetric antiresonances.
The magnitude and the width of the persistent current peak in
the vicinity of antiresonances depends on the strength of the
imaginary part of the pole. If the two poles have different
imaginary parts, the peak value of the persistent current will
be higher (along with smaller width) for the persistent current
behavior near the pole with smaller imaginary part as compared
to the larger one.

We have shown above that the persistent currents can arise in
absence of magnetic field in a open loop connected to two
reservoirs in the presence of a transport current. For fixed
value of Fermi energy the persistent currents changes sign as we
change the direction of the current flow. In equilibrium
(i.e., $\mu_{1}=\mu_{2}$) we do not get any persistent currents
in the absence of magnetic field. In the nonequilibrium
situation (i.e., $\mu_{1} \ne \mu_{2}$) it is possible to
observe the persistent currents. If $\mu_{1} > \mu_{2}$, then at
the zero temperature the total magnitude of the persistent
current is given by $I_{T}=\int _{\mu_{1}}^{\mu_{2}} I_{c}dE$.
Experimentally it is possible to observe these currents if one
tunes the Fermi energy around the antiresonances in the two port
conductance (or transmission coefficient). Moreover it is better
to tune the Fermi energy around the symmetric antiresonance so
that at finite temperature the effect survives, i.e., the
current on both sides of this antiresonance has same sign and
hence finite temperature does not lead to cancellations
as against the case of Fermi energy around asymmetrical
antiresonances.

{\bf Acknowledgements}

One of us (AMJ) thanks Professors A. G. Aronov, V. E. Kravtsov
and N. Kumar for several useful discussions on the subject of
persistent currents in open systems. Prof. Aronov also brought
to our notice the existence of circulating currents in classical
a.c. networks. A. M. J. also thanks International Center for
Theoretical Physics, Trieste, Italy for hospitality.

\vfill
\eject


\begin{thebibliography}{99}
\bibitem{1} Quantum Coherence in Mesoscopic Systems,
ed. B. Kramer, NATO ASI series, B Vol {\bf 254}, Plenum (1991).
\bibitem{2} L. P. Levy, G. Dolan, J. Dunsmuir and
H. Bouchiat, Phys. Rev. Lett, {\bf 64}, (1990) 2074.
\bibitem{3} V. Chandrasekhar, R. A. Webb, M. J. Brady, M. B.
Ketchen, W. G. Gallagher and A. Kleinasser, Phys. Rev. Lett.
{\bf 67}, 3578(1991).
\bibitem{4} D. Mailly, C. Chapelier and A. Benoit, Phys. Rev.
Lett. {\bf 70}, 2020(1993).
\bibitem{5} M. B$\ddot u$ttiker, Y. Imry and R. Landauer, Phys.
Lett. A {\bf 96} (1983) 365.
\bibitem{6} H. Cheung, Y. Gefen, E. K. Riedel, W. Shih,
Phys. Rev. B {\bf 37} (1988) 6050.
\bibitem{7} H. F. Cheung, and E. K. Riedel, Phys. Rev. B {\bf
40}, 9498(1989).
\bibitem{8} G. Montambaux, H. Bouchiat, D. Sigeti, and R.
Friesner, Phys. Rev. B {\bf 42}, 7647(1990).
\bibitem{9} D. Loss, P. Goldbart, and A. V. Balatsky, Phys. Rev.
Lett. {\bf 65}, 1655(1990).
\bibitem{10} V. Ambegaokar and U. Eckern, Phys. Rev. Lett. {\bf
65}, 381(1990).
\bibitem{11} M. Abraham and R. Berkovits, Phys. Rev. Lett. {\bf
70}, 1509(1993).
\bibitem{12} O. Entin-Wohlman, Y. Geffen, Y. Meier, and Y. Oreg,
Phys. Rev. B {\bf 45}, 11890(1992).
\bibitem{13} M. B$\ddot u$ttiker,
Phys. Rev. B {\bf 32}, 1846(1985).
\bibitem{14} M. B$\ddot u$ttiker, Squid '85- Superconducting quantum
interference devices and their application, ed H. D. Hahlbohm
and L$\ddot u$ebbig, Walter de Gruyter and Co. Berlin. New York(1985).
\bibitem{15} E. Akkermans, A. Aurbach, J. E. Avron, and B.
Shapiro, Phys. Rev. Lett. {\bf 66}, 76(1991).
\bibitem{16} P. A. Mello, Phys. Rev. B {\bf 47}, 16358(1993).
\bibitem{17} P. Singha Deo and A. M. Jayannavar,
Mod. Phys. Lett. B {\bf 7} 1045(1993).
\bibitem{18} A. M. Jayannavar and P. Singha Deo, Phys. Rev. B
{\bf 49}, 13685(1994).
\bibitem{18a} D. Takai and K. Ohta, Phys. Rev. B {\bf 48},
14318(1993).
\bibitem{19} D. F. Shaw, An introduction to electronics (second
edition), Longman, London. (1970), page 51.
\bibitem{20} A. M. Jayannavar and P. Singha Deo,
Mod. Phys. Lett. B. {\bf 5} 301(1994).
\bibitem{21} P. Singha Deo and A. M. Jayannavar Phys. Rev. B
(Oct 1994) in print.
\bibitem{22} J. Xia, Phys. Rev. B {\bf 45}, 3593(1992).
\bibitem{24} D. Takai and K. Ohta, Phys. Rev. B. {\bf 50},
2685(1994).
\bibitem{25} Y. Gefen, Y. Imry, M. Ya Azbel, Phys. Rev. Lett.
{\bf 52}, 129(1984).
\bibitem{23} E. Tekman and P. F. Bagwell, Phys. Rev. B
{\bf 48}, 2553(1993).

\end{thebibliography}
\end{document}